\documentstyle[prl,aps,preprint]{revtex}
\begin{document}
\title{Superfluid Phase Transitions in Dense Neutron Matter}
\author{V. A. Khodel,$^{1,2}$ J. W. Clark,$^1$ and M. V. Zverev$^2$}
\address{$^1$McDonnell Center for the Space Sciences
and Department of Physics,\\
Washington University, St. Louis, MO 63130 USA}
\address{$^2$Russian Research Center Kurchatov Institute,
Moscow 123182, Russia}
\date{\today , Submitted to Physical Review Letters}\maketitle

\begin{abstract}
The phase transitions in a realistic system with triplet pairing, dense
neutron matter, have been investigated.  The spectrum of phases of
the $^3P_2$--$^3F_2$ model, which adequately describes pairing in
this system, is analytically constructed with the aid of a separation
method for solving BCS gap equations in states of arbitrary angular
momentum.  In addition to solutions involving a single value of the magnetic
quantum number (and its negative), there exist ten real multicomponent
solutions.  Five of the corresponding angle-dependent order parameters
have nodes, and five do not.  In contrast to the case of superfluid
$^3$He, transitions occur between phases with nodeless order parameters.
The temperature dependence of the competition between the various
phases is studied.

\end{abstract}

\vskip 1cm

Superfluid systems manifesting triplet pairing between constituent
spin-1/2 fermions are exemplified by liquid $^3$He at millikelvin
temperatures and by the neutronic component of the quantum fluid interior
of a neutron star.  In the latter case, the density becomes high enough
that the familiar singlet $S$-wave gap has closed and pairing
in the $^3P_2$ channel is favored by the spin-orbit force acting
between neutrons \cite{hof,tam,takt,ost}. The determination of superfluid
phase diagrams for such systems over an extensive temperature range has
presented a difficult challenge for theorists.  On the one hand, the
standard Ginzburg-Landau approach is valid only for temperatures near
the critical temperature $T_c$.  On the other hand, iterative procedures,
commonly used to calculate the energy gap in systems with $S$-wave pairing,
suffer from slow convergence and uncertain accuracy when applied to the
many coupled nonlinear integral equations that arise for BCS pairing
in higher angular momentum states.

Limitations of the standard iterative approaches are accentuated when
one attempts to construct the superfluid phase diagram of the system,
which is orchestrated by tiny energy splittings between the different
solutions of the BCS pairing problem.  This problem is ordinarily
framed in terms of the set of equations \cite{ost}
\begin{eqnarray}
 \Delta_L^{JM}(p)
&=&\sum_{L'L_1J_1M_1}(-1)^{1+{L-L'\over 2}}\int \int
  \langle p| V_{LL'}^J |p_1 \rangle S^{JMJ_1M_1}_{L'L_1}({\bf n}_1)
  \nonumber  \\
&\qquad& \times
  {\tanh \left(E({\bf p}_1) / 2T\right)\over 2 E({\bf p}_1)}
  \Delta_{L_1}^{J_1M_1}(p_1) p^2_1 dp_1d{\bf n}_1
\label{bc}
\end{eqnarray}
for the coupled partial-wave components that appear in the expansion\hfill\break
$\Delta_{\alpha\beta}({\bf p}) =\sum_{J,L,M}\Delta_L^{JM}(p)
\left(G_{LJ}^M({\bf n})\right)_{\alpha\beta}$ of the $2 \times 2$
gap matrix in terms of the spin-angle matrices
$\left(G_{LJ}^M({\bf n})\right)_{\alpha\beta}=\sum_{M_SM_L}
C^{1M_S}_{{1\over 2}{1\over 2}\alpha\beta}C^{JM}_{1LM_SM_L}
Y_{LM_L}({\bf n})$.  The quasiparticle energy
\begin{equation}
E({\bf p})=\left[\xi^2(p)+{1\over 2}\sum_{LJML_1J_1M_1}
\left(\Delta_L^{JM}(p)\right)^* \Delta_{L_1}^{J_1M_1}(p)
S^{JMJ_1M_1}_{LL_1}({\bf n})\right]^{1\over 2} \, ,
\end{equation}
is constructed from the gap components $\Delta_L^{JM}(p)$ together
with the single-particle spectrum $\xi(p)\simeq p_F(p-p_F)/M^*$
of the normal Fermi liquid, where $M^*$ is a suitable effective
mass.  This quantity is rendered angle dependent by the spin trace
$S_{LL_1}^{JMJ_1M_1}({\bf n}) = {\rm Tr}
\left[\left(G_L^{JM}({\bf n})\right)^*G_{L_1}^{J_1M_1}({\bf n})\right]$,
further complicating explicit solution of the system (1).

The pairing matrix elements $\langle p| V_{LL'}^J|p_1 \rangle$ are
generated by the spin-angle expansion $V({\bf  p},{\bf p}_1)
=\sum_{LL'JM}(-1)^{{L-L'\over 2}} \langle p| V_{LL'}^J|p_1 \rangle
G_{LJ}^M({\bf n})\left(G_{L'J}^M({\bf n}_1)\right)^*$
of the block of Feynman diagrams irreducible in the particle-particle
channel.  A salient feature of the vacuum $nn$ interaction is that the
components of the central forces nearly compensate each other, as
evidenced in the experimental $P$-scattering phases \cite{arndt}.
We shall assume that this feature is preserved by the effective $nn$
interaction inside neutron matter.  The dominant role of the spin-orbit
component in promoting the $^3P_2$ pairing channel at super-nuclear
densities then implies that contributions to triplet pairing from
nondiagonal terms with $L',\, L_1\neq 1$ or $J_1\neq 2$ on the
r.h.s.\ of Eq.~(1) can be evaluated within perturbation theory,
in terms of the set of principal gap amplitudes $\Delta_1^{2M}(p)$,
with $M$ running from $-2$ to 2.  In fact, time-reversal invariance
implies that the problem may be treated in terms of only three complex
functions, namely $\Delta_1^{2M}(p)$ with $M=0\,,1\,,2$.

Essential further simplifications are possible if we apply the separation
approach developed in \cite{kkc2}, which establishes the factorization
\begin{equation}
\Delta^{2M}_1(p)=D^{2M}_1\chi(p)\  ,\qquad M=0,1,2 \,  ,
\end{equation}
where $\chi(p)$ is a universal shape factor normalized to unity
at $p=p_F$.  As a consequence, the elucidation of the
phase diagram of dense superfluid neutron matter reduces to the
determination of the three coefficients $D_1^{2M}$, since the
character of the phase diagram itself is independent of the
shape factor $\chi(p)$.

Among the nondiagonal contributions to the r.h.s.\ of Eq.~(1), two
assume leading importance.  The first contains the integral of
the product $ V_{31}^2 S^{2M2M_1}_{31}\Delta^{2M_1}_1$, while
the second contains the integral of the product
$ V_{11}^2S^{2M2M_1}_{13}\Delta^{2M_1}_3$.  Considering
only these contributions, we obtain the $^3P_2$--$^3F_2$ pairing
problem, which has been treated numerically in earlier work,
notably Refs.~\cite{takt,ost,bal,kkc3}.  The list of relevant
states appears to be exhausted with the addition of the (less
important) $^3P_0$ and $^3P_1$ pairing channels \cite{kkc3}.
The rapid convergence of the nondiagonal integrals on the r.h.s.\ of
the gap equations (1) greatly facilitates implementation of the
perturbation strategy, since the overwhelming contributions to these
integrals come from the region adjacent to the Fermi surface.  For
$E({\bf p})$ significantly in excess of the energy gap value $\Delta_F$,
the energies $E({\bf p})$ and $|\xi(p)|$ practically coincide and, as a
result, the angular integration yields zero.  Thus, when dealing with
the nondiagonal contributions, it is sufficient to know the minor gap
components $\Delta_3^{2M}(p)$ (with $M=0\,,1\,,2$) at the point $p=p_F$,
which may be efficiently evaluated in terms of the coefficients $D^{2M}_1$
with the aid of the set (\ref{bc}) itself.  In doing so, we retain
on the r.h.s.\ of (\ref{bc}) only the dominant contribution containing
a large logarithmic factor $\ln (\epsilon_F/\Delta_F)$, where $\epsilon_F$
is the Fermi energy.  This leads to the simple connection
\begin{equation}
\Delta^{2M}_3(p=p_F)=\eta D^{2M}_1 \  ,
\label{prop1}
\end{equation}
where $\eta=-\langle p_F| V_{13}^2|p_F \rangle/v_F$
and $v_F\equiv\langle p_F| V_{11}^2|p_F \rangle$.
Analogous linear relations hold for other minor components $\Delta_L^{JM}$
of the gap function (notably $\Delta_1^{00}$ and $\Delta_1^{1M}$).

When calculated for in-vacuum neutron-neutron interactions
\cite{arg}, the $\eta$ value depends smoothly on $\rho$,
varying around $0.3$ in the interval $\rho_0<\rho<3\rho_0$ \cite{kkc3}.
In view of its relativistic origin, the spin-orbit force should not be
much affected by polarization or correlation corrections, a judgment
supported by the empirical analysis of spin-orbit splitting in
finite nuclei.  Medium modification of the tensor force may be more
significant, especially in the environs of the pion-condensation phase
transition \cite{mig,akmal}.  Investigation of this possibility calls
for special treatment and will be pursued elsewhere.

A complete understanding of the phase diagram of systems that exhibit
triplet pairing would require the inclusion of corrections to the
in-medium interaction depending on the pairing gap itself.
These ``strong-coupling'' corrections are most important near the
critical temperature $T_c$ \cite{wol,ss}.  Their effect is not addressed
in the present work, which focuses on temperatures far enough below
$T_c$ that pairing corrections to the vertex $V$ (and hence to the
parameter $\eta$) may be ignored.

We turn now to the determination of the key coefficients
$D^{2M}_1$ $(M=0,\,1,\,2)$ to leading perturbative order in the
parameter $\eta$.  Making use of the connection (\ref{prop1}), simple
manipulations of Eqs.~(1) at $p=p_F$ yield three coupled equations for
these coefficients,
\begin{equation}
D^{2M}_1+v_F\sum_{M_1}D^{2M}_1\int\int\phi(p)
{\tanh{\left( E_0({\bf p})/ 2T \right)}\over
2E_0({\bf p})} S^{2M2M_1}_{11}({\bf n})\chi(p)p^2 dp
d{\bf n}= \eta v_F r_M \, ,
\label{cutr}
\end{equation}
with $\phi(p)\equiv\langle p| V^2_{11}|p_F \rangle/v_F$,
$E_0({\bf p}) \equiv E({\bf p}; \eta=0)$, and
\begin{equation}
r_M =\sum_{M_1}D_1^{2M_1}\int\int
\left[S^{2M2M_1}_{31}({\bf n})+S^{2M2M_1}_{13}({\bf n})\right]
{\tanh{\left(E_0({\bf p})/2T\right)}\over 2E_0({\bf p})}
  p^2 dp d{\bf n} \, .
\label{rhsr}
\end{equation}

The search for solutions will be confined to those with real coefficients
$D^{2M}_1$ since, as a rule, such states lie lower in energy than
those with complex $D^{2M}_1$.  Inserting the explicit form of
$S^{2M2M_1}_{11}({\bf n})$ into Eqs.~(\ref{cutr}) and (\ref{rhsr}),
we may arrive at a system of three equations
\begin{eqnarray}
  \lambda_2+ v_F\left[\lambda_2(J_0+J_5) -\lambda_1 J_1 -J_3\right]&=&
\eta v_F r_2\, ,\nonumber \\
 \lambda_1+ v_F\left[-(\lambda_2+1)J_1+\lambda_1(J_0+4J_5+2J_3)/4\right]&=&
\eta v_F r_1\, ,\nonumber\\
1+ v_F\left[-(\lambda_2 J_3+\lambda_1 J_1)/3 +J_5\right]&=&\eta v_F r_0 \, ,
\label{sprl}
\end{eqnarray}
for the two ratios $\lambda_1=D^{21}_1/D^{20}_1\sqrt{6}$ and
$\lambda_2=D^{22}_1/D^{20}_1\sqrt{6}$ and the gap value $\Delta_F$.
These equations are written so as to coincide at $\eta=0$ with the
equations of the pure $^3P_2$ pairing model solved in
Ref.~\cite{kkc3}, where
\begin{equation}
J_i = \int\int f_i(\theta,\varphi)
\phi(p){\tanh{E_0({\bf p})/ 2T}\over 2E_0({\bf p})}\chi(p)
{p^2dpd{\bf n}\over 4\pi}
\label{intk}
\end{equation}
with $f_0=1-3z^2$, $f_1=3xz/2$, $f_3= 3(2x^2+z^2-1)/2$, and $f_5=(1+3z^2)/2 $
and $z=\cos\theta$, $x=\sin\theta\cos\phi$, and $y=\sin\theta\sin\phi$.
Of the integrals $J_i$ (with $j=1,\cdots 5$), only $J_5$ contains a
principal term going like $\ln(\epsilon_F/\Delta_F)$.

Eqs.~(\ref{sprl}) have three familiar {\it one-component} solutions
\cite{tam,ost,kkc3} corresponding to $M=0$, 1, and 2.  To establish the
structure and the spectrum of the {\it multicomponent} solutions of the
perturbed problem, we carry out a two-step transformation of the set
(\ref{sprl}).  The integral $J_5$ is responsible for introducing
the gap value $\Delta_F$ but is irrelevant to the phase structure.
Thus, as a first step we combine the equations (\ref{sprl}) so as to
eliminate terms involving $J_5$ from the first pair and, in addition,
reduce the number of the $J_i$ integrals in each equation to two.
The resulting $J_5$--independent equations are
\begin{eqnarray}
(\lambda_2+1)[  3\lambda_1(\lambda_2+1)J_0 -
           2(\lambda^2_1-2\lambda_2^2+6) J_1] &=&\eta B_1
   \, ,\nonumber \\
(\lambda_2+1)[ (\lambda_1^2-4\lambda_2)J_1 +
                    \lambda_1(\lambda_2+1)J_3] &=&\eta B_2 \  ,
\label{sp2}
\end{eqnarray}
where
$B_1=2\lambda_1(2\lambda_2+3)r_2-4(\lambda^2_2-3)r_1-
6\lambda_1(\lambda_2+2)r_0 $ and
$B_2=-\lambda_1 r_2+4\lambda_2r_1-3\lambda_1\lambda_2 r_0 $.
The second step is taken under the assumption that
$\lambda_2\neq 1$; otherwise the ensuing manipulations lose
their meaning.  As in Ref.~\cite{kkc2}, we perform the rotation
$(x,z)=(-t\sin\vartheta+u\cos\vartheta,\, t\cos\vartheta+u\sin\vartheta)$,
choosing the angle $\vartheta$ to remove the integral $J_1$
from Eqs.~(\ref{sp2}). In particular, if $\zeta=\tan\vartheta$ is
chosen to obey the algebraic equation
$\lambda_1\zeta^2-(\lambda_2-3)\zeta-\lambda_1=0$,
substitution of the transformed integrals $J_i$ into
Eqs.~(\ref{sp2}) yields
\begin{eqnarray}
(\lambda_2+1)[A_1J_0+A_2J_3]&=&\eta B_1 \ , \nonumber \\
(\lambda_2+1)[A_1J_0+A_2J_3]&=&-2 \eta B_2  \  ,
\label{eqaa}
\end{eqnarray}
with
$A_1= {3\over
2}\lambda_1(1+\lambda_2)(2-\zeta^2)
-{3\over 2}(\lambda_1^2-2\lambda_2^2+6)\zeta$ and
$A_2=-3\lambda_1(1+\lambda_2)\zeta^2-(\lambda_1^2 -2\lambda_2^2+6)\zeta$.

We observe that the left-hand members of these two equations (12) are
identical.  It is just this feature that leads to the universalities
of the pure $^3P_2$ pairing problem discovered in Ref.~\cite{kkc2}.
Independently of temperature, density, and details of the
in-medium interaction, the solutions of this restricted problem
derived from Eqs.~(9) at $\eta\equiv 0$ (where the right-hand
members of Eqs.~(12) are trivially coincident) fall into two groups
degenerate in energy, namely an upper group comprised of
states whose angle-dependent order parameters have nodes and
a lower group with nodeless order parameters (cf.\ Ref.~\cite{rich}).
In addition to the energy degeneracies, the multicomponent pairing
solutions, which satisfy
\begin{equation}
(\lambda^2_1+2-2\lambda_2)(\lambda^2_1-2\lambda^2_2-6\lambda_2)=0  \, ,
\label{purep}
\end{equation}
display a degeneracy with respect to the coefficient ratios
$\lambda_1$ and $\lambda_2$, since they generally define {\it curves}
rather than {\it points} in the $(\lambda_1,\lambda_2)$ plane.

The strong parametric degeneracy inherent in pure $^3P_2$
pairing is lifted in the case of $^3P_2$--$^3F_2$ pairing: at $\eta > 0$
the true solutions of the problem are represented by a set of
isolated points in the $(\lambda_1,\lambda_2)$ plane. Indeed, upon
equating the right-hand members of Eqs.~(\ref{eqaa}) in the small--$\eta$
limit, one obtains an additional relation between the parameters
$\lambda_1(\eta=0)$ and $\lambda_2(\eta=0)$, viz.
\begin{equation}
\lambda_1 r_2-(\lambda_2-3)r_1-3\lambda_1 r_0=0 \  ,
\label{rat2}
\end{equation}
where the $r_M$ are defined by Eq.~(\ref{rhsr}).  This relation
supplements the spectral condition (11) and lifts the parametric
degeneracy.

The system formed by (\ref{purep}) and (\ref{rat2}) is solved
analytically by utilizing the same rotation in $x-z$ coordinates as
described above.   After lengthy algebra, one arrives at the full
set of solutions of the coupled-channel $^3P_2$--$^3F_2$ pairing
problem.  A conspicuous feature of these solutions, made transparent
by the separation method \cite{kkc2}, is their virtually complete
independence of the temperature $T$.

Before cataloging and classifying the $^3P_2$--$^3F_2$ pairing
solutions, it is worth pointing out that the particular solution
$\lambda_2=-1$ found in the pure $^3P_2$ problem remains intact
upon switching on the $^3F_2$ channel, but it completely disappears
if any further channel (e.g. $^3P_0$) enters the picture.

Fig.~1 summarizes the results of our analytical considerations.
Only the right half of the $(\lambda_1,\lambda_2)$ plane is shown,
since the relevant energies are independent of the sign of $\lambda_1$.
Besides the well-known single-component solutions with $|M|=0$, 1,
or 2, the collection of unitary solutions of the $^3P_2$--$^3F_2$
pairing problem contains {\it ten} multicomponent solutions, corresponding
to more complicated superfluid phases.  {\it Five} of these additional
solutions, denoted $O_k$ ($k=1,\ldots,5$), have nodeless order parameters
and include:
\begin{itemize}
\item[(a)]
{\it Two} two-component solutions $O_{\pm3}$, identical to those found in the
pure $^3P_2$ pairing problem, with $\lambda_1=0$ and $\lambda_2=\pm
3$.
\item[(b)]
{\it Three} three-component solutions, two of which, $O_1$ and $O_4$,
are associated with the upper branch of
$\lambda^2_1-2\lambda^2_2-6\lambda_2=0$ and have
$\lambda_2=3(\sqrt{21}{-}4)/5$ and $\lambda_2=3$, respectively;
while a third, $O_2$, is associated with the lower branch of the
same equation and has $\lambda_2=-3(\sqrt{21}+4)/5$.
\end{itemize}
The remaining five solutions of the problem, denoted by $X_k$ ($k=1,
\ldots,5$), are distinguished by order parameters that do have
zeros.  This set includes:
\begin{itemize}
\item[(a)]
{\it Two} two-component solutions $X_{\pm1}$, again identical to
those found in the pure $^3P_2$ pairing problem, with
$\lambda_1=0$ and $\lambda_2=\pm 1$.
\item[(b)]
{\it Three} three-component solutions, $X_2$, $X_3$, and $X_4$,
associated with the parabola $\lambda_2=\lambda^2_1/2+1$ and
having $\lambda_2=13-2\sqrt{35}$, $\lambda_2=3$, and
$\lambda_2=13+2\sqrt{35}$, respectively.
\end{itemize}
These general features of the spectrum of solutions of the
$^3P_2$--$^3F_2$ problem are expected to persist even if $\eta$
is not so small.

To construct the phase diagram of superfluid neutron matter,
the third of Eqs.~(\ref{sprl}) must be brought into play
to determine the gap values $\Delta_F$ for the different
solutions.  We must then compare the free-energy shifts
$F_s=-\int \Delta^2_F(g)g^{-2}dg$ due to pairing
in the corresponding superfluid states, where $g$ is the
relevant pairing coupling constant.

At low $T$, the only true contestants for existence and ascendancy
are the solutions with nodeless order parameters, since the other
solutions lie too high in energy; the gap between the states of the two
different groups cannot be bridged if the value of $|\eta|$ stays
rather small.  Thus, even though the degeneracy of the $^3P_2$ pairing
problem in the $\lambda_1-\lambda_2$ plane, embodied in the relation
(11), is entirely removed when the $^3P_2$--$^3F_2$ coupling is switched
on, the degeneracies in the energetic spectrum of the different
superfluid phases are only partially lifted.  Instead, this spectrum
decays into several groups of nearly degenerate states, the $O_1$ and
$O_2$ phases forming the lowest-energy group, the next higher group
being composed of the phases $O_{\pm 3}$ along with the one-component
phase with $M=0$, and so on.  As seen in Fig.~2, the splitting
between the two groups lowest in energy shrinks as $T$ increases,
until, at $T\simeq 0.7T_c$, their roles are interchanged and
transitions occur.  However, unlike the A-B phase transition in
superfluid $^3$He, where the order parameter of the A-phase solution
has nodes while that of the B-phase does not, the transition in neutron
matter takes place between phases with nodeless order parameters.
Introduction of the $^3P_0$ and $^3P_1$ pairing channels produces
a further lifting of energy degeneracies and hence a further
complication of the phase diagram of superfluid neutron matter.
Unfortunately, it is not possible to resolve the resulting fine
structure of the phase spectrum without more detailed information
on the in-medium effective interaction between neutrons, which
is currently unavailable.

The temperature region near $T_c$ provides another venue for possible
superfluid phase transitions.  In this regime, strong-coupling
corrections should no longer be ignored \cite{ss}, and the character
of the phase diagram may be influenced significantly by external
magnetic fields.  Finally, fermion condensation, occurring as a
precursor to pion condensation \cite{vosk}, should be mentioned as
a source of exotic phase transitions in superfluid neuron matter at
densities approaching the critical value for collapse of the pion
mode \cite{mig}.

If, as we have shown, the phase diagram of dense neutron matter
exhibits several triplet superfluid phases, then phase transitions
between different phases are expected to occur as the neutron
star cools.  Since the gap value changes in these transitions, their
occurrence may ultimately be detected in the thermal history and/or
the rotational dynamics of the star, for example in variation of
its moment of inertia or in alterations of the distribution of the
angular momentum between the crust and the vortex system, resulting
in a change of the star's angular velocity.  Possible scenarios
leading to distinctive observable manifestations will be presented
separately.

We thank A.~ D.~Sedrakian, I.~I.~Strakovsky, G.~E.~Volovik,
and D.~N.~Voskresensky for illuminating discussions.  This research
was supported in part by the U. S. National Science Foundation under
Grant No.~PHY-9900713 (JWC and VAK), by the McDonnell Center for
the Space Sciences (VAK), and by Grant No.~00-15-96590 from the
Russian Foundation for Basic Research (VAK and MVZ).  MVZ acknowledges
the hospitality of the INFN (Sezione di Catania).

\newpage
\centerline{\bf Figure Captions}

\vskip 0.5 cm
\noindent
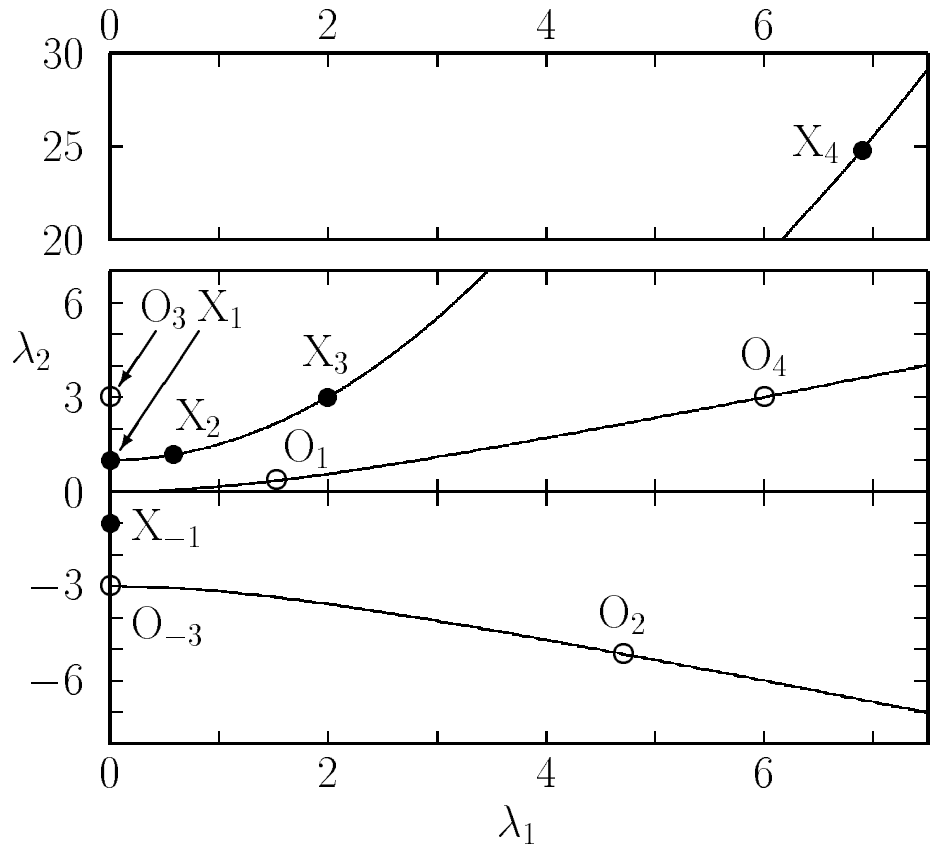
{\bf Fig.~1.}  Multicomponent solutions of the $^3P_2$--$^3F_2$
pairing problem.  Solutions whose order parameter is nodeless [exhibit nodes]
in the case of pure $^3P_2$ pairing are indicated by filled [open]
circles.

\vskip 0.2 cm
\noindent
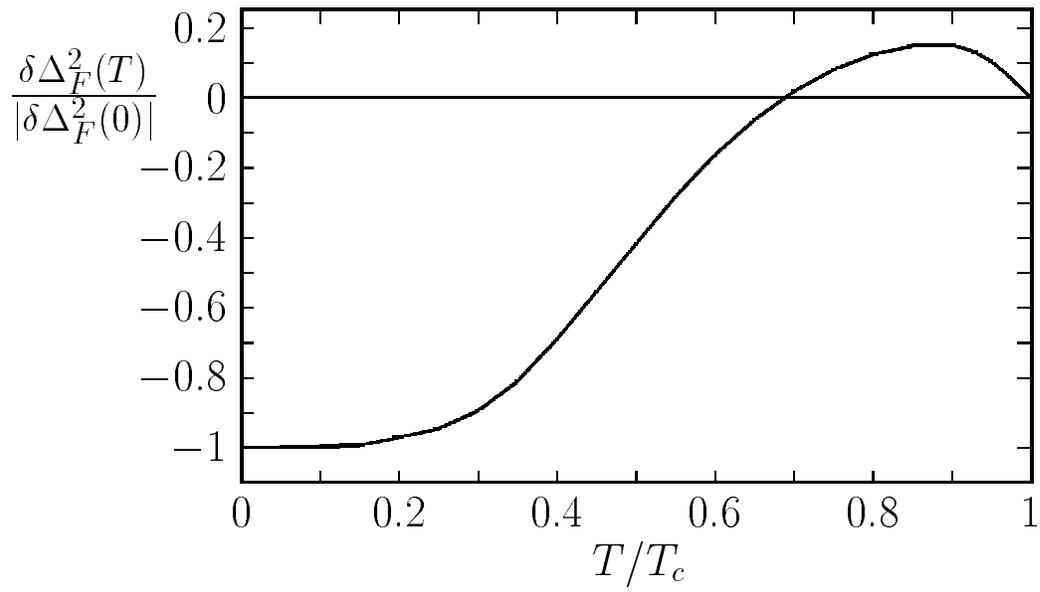
{\bf Fig.~2.}
Difference $\delta\Delta^2_F(T)$ between the $\Delta^2_F(T)$ values for
the phases $O_1$ or $O_2$ and for the phases $O_{\pm 3}$ (or the
one-component phase with $M=0$), measured in units of
$|\delta\Delta^2_F(T=0)|$.
\end{document}